\documentstyle[11pt]{article}\textheight 230mm\textwidth 150mm
            \newcommand{\be}{\begin{eqnarray}}
            \newcommand{\ee}{\end{eqnarray}}
            \newcommand{\eel}[1]{\label{#1}\end{eqnarray}}
\newcommand{\e}[1]{\label{e:#1}\end{eqnarray}}
     \newcommand{\eg}{{\em e.g.\ }}
            \newcommand{\ie}{{\em i.e.\ }}

            \newcommand{\la}{{\lambda}}
            
            \newcommand{\del}{{\delta}}

           \newcommand{\ra}{{\rightarrow}}
\newcommand{\Ra}{{\Rightarrow}}
 \newcommand{\lea}{{\leftarrow}}
            
            \newcommand{\Lra}{{\Leftrightarrow}}
            \newcommand{\pet}{{\cal P}}

\newcommand{\ca}{{\cal C}}

            \newcommand{\beq}{\begin{quote}}
            \newcommand{\eq}{\end{quote}}
            \newcommand{\Om}{\Omega}

            \newcommand{\ben}{\begin{enumerate}}
            \newcommand{\een}{\end{enumerate}}
            \newcommand{\bit}{\begin{itemize}}
            \newcommand{\ei}{\end{itemize}}
    	\newcommand{\nn}{\nonumber}
            \newcommand{\r}[1]{(\ref{e:#1})}
            \newcommand{\edfl}[1]{\label{#1}\end{df}}

\newcommand{\ve}{{\varepsilon}}

\newcommand{\dagg}{^{\dag}}

\def\d{\partial}
\def\cC{{\cal C}}

  \def\half{{1 \over 2}}

\begin{document}
\begin{titlepage}
\noindent
G\"{o}teborg ITP 98-07\\
(Revised version) \\

\vspace*{5 mm}
\vspace*{35mm}
\begin{center}{\LARGE\bf Open groups of constraints\\ - Integrating arbitrary
involutions -}
\end{center} \vspace*{3 mm} \begin{center} \vspace*{3 mm}

\begin{center}Igor Batalin\footnote{On leave of absence from
P.N.Lebedev Physical Institute, 117924  Moscow, Russia\\E-mail:
batalin@td.lpi.ac.ru.} and Robert
Marnelius\footnote{E-mail: tferm@fy.chalmers.se}\\ \vspace*{7 mm} {\sl
Institute of Theoretical Physics\\
Chalmers University of Technology\\ G\"{o}teborg University\\
S-412 96  G\"{o}teborg, Sweden}\end{center}
\vspace*{25 mm}
\begin{abstract}
A new type of quantum master equation is presented which is expressed in
terms of a
recently introduced quantum antibracket. The equation involves only two
operators:
an extended nilpotent BFV-BRST charge and an extended ghost charge. It
is proposed to determine the generalized quantum Maurer-Cartan equations for
arbitrary open groups. These groups are the integration of constraints in
arbitrary
involutions. The only condition for this is that the constraint operators may be
embedded in an odd nilpotent operator,  the BFV-BRST charge. The proposal is
verified at the quasigroup level. The integration formulas are also used to
 construct a generating operator for quantum antibrackets of operators in
arbitrary involutions.
\end{abstract}\end{center}\end{titlepage}

\setcounter{page}{1}
\setcounter{equation}{0}
\section{Introduction.}
In \cite{Quanti} we have introduced quantum antibrackets. They are new objects
which cast new light on the BV-quantization of arbitrary gauge theories. Here we
show that they are also useful in a more general context. In terms of the
quantum
antibracket we give a new simple quantum master equation which we conjecture
to encode the Maurer-Cartan equations and their generalizations to open  groups
whose generators are in arbitrary involutions. This conjecture is verified
at the
quasigroup level. The new master equation involves only two operators: an
extended
nilpotent BFV-BRST charge and an extended ghost charge. The construction is thus
embedded in a general BRST framework which should make the formulas useful for
general gauge theories.

In section 2 we review the BFV-treatment of constraints in arbitrary
involutions,
and in section 3 we present the new master equation for generalized
quantum Maurer-Cartan equations. In section 4 we treat quasigroup theories in
detail, and in section 5  the formalism is used to generalize the generating
operator  for quantum antibrackets given in \cite{Quanti} to operators in
arbitrary
involutions.
\section{Constraints in arbitrary involutions.}
Consider classical, real constraints $\theta_a$ on a symplectic manifold. Their
Grassmann parities are
$\ve(\theta_a)\equiv\ve_a (=0, 1)$. $\theta_a$ are assumed to be in arbitrary
involutions with respect to the Poisson bracket, \ie
 \be
 &&\{\theta_a, \theta_b\}=U_{ab}^{\;\;\;c}\theta_c,
\e{1}
where the structure coefficients $U_{ab}^{\;\;\;c}$ may be arbitrary  functions
on the considered symplectic manifold. It is well-known that this algebra always
may be embedded in one single real, odd function $\Om$ on a ghost extended
manifold
in such a way that $\{\Om, \Om\}=0$ in terms of the extended Poisson
bracket. $\Om$
is the BFV-BRST charge \cite{BFV}. The corresponding quantum theory
is consistent if the corresponding odd, hermitian operator
$\Om$ is nilpotent, \ie $\Om^2=0$. For a finite number of degrees of
freedom such a
solution always exists and is of the form \cite{BF} ($N$ is the rank of the
theory
and $\theta_a$ are the hermitian constraint operators)
\be
 &&\Om=\sum_{i=0}^N\Om_i,
 \e{2}
\be
&&\Om_0\equiv\ca^a\theta_a,\;\;\;\Om_i\equiv \Om_{a_1\cdots a_{i+1}}^{b_i\cdots
b_1}(\pet_{b_1}\cdots\pet_{b_i}\ca^{a_{i+1}}\cdots\ca^{a_1})_{Weyl},
\;\;\;i=1,\ldots,N,
\e{3}
where we have introduced the ghost operators $\cC^a$, $\pet_a$,
$\ve(\cC^a)=\ve_a+1$, satisfying
 \be
&&[\ca^a,
\pet_b]=i\hbar\del^a_b,\quad(\cC^a)\dagg=\cC^a,
\quad\pet_a\dagg=-(-1)^{\varepsilon_a}\pet_a.
\e{4}
In \r{3} the ghost operators are Weyl ordered which means that $\Om_i$ are all
hermitian.
One may notice that $\Om$ in \r{2} has ghost number one, \ie
\be
&&[G, \Om]=i\hbar\Om,\quad
G\equiv-\half\left(\pet_a\cC^a-\cC^a\pet_a(-1)^{\ve_a}\right),
\e{5}
where $G$ is the ghost charge. $\Om$ determines  the
precise form of the quantum counterpart of the algebra \r{1}. A convenient
form of
this algebra is obtained if we rewrite $\Om$ in the
 following $\cC\pet$-ordered form
\cite{BF}
 \be
 &&\Om=\sum_{i=0}^N\Om'_i,\;\;\;\Om'_0\equiv\ca^a\theta'_a,\nn\\
&&\Om'_i\equiv \ca^{a_{i+1}}\cdots\ca^{a_1}{\Om'}_{a_1\cdots a_{i+1}}^{b_i\cdots
b_1}\pet_{b_1}\cdots\pet_{b_i},\;\;\;i=1,\ldots,N.
 \e{51}
The  nilpotency of $\Om$ requires then the algebra
\be
&&[\theta'_a, \theta'_b]=i\hbar{U'}_{ab}^{\;\;\;c}\theta'_c,
\e{6}
where the structure operators
${U'}_{ab}^{\;\;\;c}$ are given by
\be
&&{U'}_{ab}^{\;\;\;c}=2(-1)^{\ve_b+\ve_c}{\Om'}_{ab}^{\;\;\;c}.
\e{7}
In terms of the coefficient operators in \r{3} ${\Om'}_{ab}^{\;\;\;c}$ and
$\theta'_a$ are given by
\be
&&{\Om'}_{ab}^c={\Om}_{ab}^c
+\half\sum_{n=1}^\infty\left({i\hbar\over 2}\right)^n(n+1)(n+2)!
\,\Om_{aba_1\cdots
a_n}^{a_n\cdots a_1 c}(-1)^{\sum_{k=1}^n\ve_{a_k}},\nn\\
&&\theta_a'=\theta_a+\sum_{n=1}^\infty\left({i\hbar\over
2}\right)^n(n+1)!\,\Om^{a_n\cdots a_1}_{aa_1\cdots a_n}
(-1)^{\sum_{k=1}^n\ve_{a_k}},
\e{71}
which shows that  $\theta'_a$  in general are different from $\theta_a$ and  in
general not even hermitian. However, the main point here is that
$\Om$  through \eg
\r{6} represents the quantum counterpart of
\r{1}.

\section{Quantum master equation and generalized Maurer-Cartan equations.}
We want now to integrate the quantum involution \r{1} encoded in $\Om$ as
represented by \eg \r{6}, or in other words we are looking for finite gauge
transformations if $\theta'_a$ are viewed as gauge generators.  We consider
therefore the  Lie equations
\be
&&A(\phi)\stackrel{\lea}{\nabla}_a\equiv
A(\phi)\stackrel{\lea}{\d_a}-(i\hbar)^{-1}[A(\phi), Y_a(\phi)]=0,
\e{8}
where $\d_a$ is a derivative with respect to the parameter $\phi^a$,
$\ve(\phi^a)=\ve_a$. The operator $Y_a$, which depends on $\phi^a$, must
satisfy integrability conditions
\be
Y_a\stackrel{\lea}{\d_b}-Y_b\stackrel{\lea}{\d_a}(-1)^{\ve_a\ve_b}=
(i\hbar)^{-1}[Y_a,
Y_b],
\e{9}
which in turn are integrable without further conditions. In order for the
Lie equations \r{8} to be connected to the integration of the quantum involution
\r{1} $Y_a(\phi)$ has to be of the form
\be
&&Y_a(\phi)=\la^b_a(\phi)\theta'_b(-1)^{\ve_a+\ve_b}+\{\mbox{\small
possible ghost
dependent terms}\},\quad
\la^b_a(0)=\del^b_a,
\e{10}
where $\la^b_a(\phi)$ are operators in general. One may note that in a ghost
independent scheme
$\theta'_a$ are the generators of the finite transformations. However,
within our
BRST framework, $[\Om, \pet_a]=\theta'_a+\{${\small ghost dependent
terms}$\}$, are
the appropriate generators. This motivates the form \r{10} and it also makes it
natural to expect
$Y_a$ to have the general form
\be
&&Y_a(\phi)=(i\hbar)^{-1}[\Om, \Om_a(\phi)],\quad\ve(\Om_a)=\ve_a+1,
\e{12}
where $Y_a$ has ghost number zero and $\Om_a$ ghost number minus one.
This form implies that $[Y_a(\phi), \Om]=0$, so that if $[A(0), \Om]=0$
then $[A(\phi), \Om]=0$, \ie a BRST invariant operator remains BRST
invariant when
 transformed according to
\r{8}. In the following we consider $Y_a$ to be given
by $\Om_a$ through equation \r{12}. Equations
\r{10} and
\r{12}  together with
\r{2} and \r{3}  imply that  $\Om_a$ in \r{12} must be of the form
\be
&&\Om_a(\phi)=\la^b_a(\phi)\pet_b+\{\mbox{\small possible ghost
dependent terms}\},\quad
\la^b_a(0)=\del^b_a.
\e{11}
The integrability condition \r{9} for $Y_a$ leads then by means of \r{12} to the
following equivalent equation for
$\Om_a$
\be
&&\Om_a\stackrel{\lea}{\d_b}-\Om_b\stackrel{\lea}{\d_a}(-1)^{\ve_a\ve_b}=
(i\hbar)^{-2}(\Om_a, \Om_b)_{\Om}-\half(i\hbar)^{-1}[\Om_{ab}, \Om],
\e{13}
where we have introduced the quantum antibracket defined in accordance with
\cite{Quanti},
\ie
\be
&&(A, B)_{Q}\equiv\half\left([A, [Q, B]]-[B, [Q,
A]](-1)^{(\ve_A+1)(\ve_B+1)}\right)=\nn\\&&=\half\left([[A, Q], B]-[[B, Q],
A](-1)^{(\ve_A+1)(\ve_B+1)}\right), \quad Q^2=0.
\e{14}
This expression satisfies all properties of an antibracket if $A$ and $B$
commute, or
for more general
$A$'s and
$B$'s  under certain conditions \cite{Quanti}. Due to the form
 \r{11} of $\Om_a$, eq.\r{13} are generalized Maurer-Cartan equations for
$\la^b_a(\phi)$. The integrability conditions of \r{13} lead to
equivalent first order equations of $\Om_{ab}$ and so on. Thus,  $Y_a$ is
replaced by a whole set of operators, and  the integrability condition \r{9} for
$Y_a$ is replaced by a whole set of integrability conditions. What is
amazing  is
that \r{13} together with all its integrability conditions seem to be
possible to
embed in one single quantum master equation (eq.\r{18} below) which
involves only two
operators. The first operator is an extended nilpotent BFV-BRST charge
involving the
conjugate momenta
$\pi_a$ to
$\phi^a$,  now  turned into operators, and
new ghost variables $\eta^a$, $\ve(\eta^a)=\ve_a+1$, to  be treated as
parameters. It is
\be
&&\Delta\equiv\Om+\eta^a \pi_a(-1)^{\ve_a},\quad \Delta^2=0,\quad
[\phi^a, \pi_b]=i\hbar\del^a_b.
\e{15}
The second operator in the master equation is an even
 operator $S(\phi, \eta)$ defined by
\be
&&S(\phi,
\eta)\equiv G+\eta^a\Om_a(\phi)+\half\eta^b\eta^a\Om_{ab}(\phi)(-1)^{\ve_b}+
\nn\\&&+{1\over6}\eta^c\eta^b\eta^a\Om_{abc}(\phi)(-1)^{\ve_b+\ve_a\ve_c}+
\ldots\nn\\&&\ldots+
{1\over n!}\eta^{a_n}\cdots\eta^{a_1}\Om_{a_1\cdots
a_n}(\phi)(-1)^{(\ve_{a_2}+\ldots+\ve_{a_{n-1}}+\ve_{a_1}\ve_{a_n})}+\ldots,
\e{16}
where $G$ is the ghost charge operator in \r{5}. Our main conjecture is that the
operators
$\Om_{a_1\cdots a_n}(\phi)$ in \r{16} may  be identified with $\Om_{a}$,
$\Om_{ab}$ in
\r{13} and all the $\Om$'s in their integrability conditions in a
particular manner.
They satisfy the properties
\be
&&\ve(\Om_{a_1\cdots a_n})=\ve_{a_1}+\ldots+\ve_{a_n}+n,\quad[G, \Om_{a_1\cdots
a_n}]=-n i\hbar\Om_{a_1\cdots a_n}.
\e{17}
The last relation implies that $\Om_{a_1\cdots a_n}$ has ghost number minus
$n$.
If we assign ghost
number one to $\eta^a$ then $\Delta$ has ghost number one and $S$ has ghost
number
zero. The operators $\Om_{a_1\cdots a_n}(\phi)$ are determined by the following
quantum master equation
\be
&&(S, S)_{\Delta}=i\hbar[\Delta, S],
\e{18}
where $(S, S)_{\Delta}$ is the quantum antibracket
defined in accordance with \r{14}. Thus, we have
\be
&&\quad (S, S)_{\Delta}\equiv[[ S, \Delta], S].
\e{181}
 Consistency requires  $[\Delta, S]$ to be
nilpotent  since
\be
&&[\Delta, (S, S)_{\Delta}]=0\quad \Lra\quad [\Delta, S]^2=0.
\e{19}
The explicit form of $[S, \Delta]$ to the lowest orders in $\eta^a$ is
\be
&&[S, \Delta]=i\hbar\Om+\eta^a[\Om_a,
\Om]+\eta^b\eta^a\Om_a\stackrel{\lea}{\d_b}i\hbar(-1)^{\ve_b}+
\half\eta^b\eta^a[\Om_{ab},
\Om](-1)^{\ve_b}+\nn\\&&+\half\eta^c\eta^b\eta^a\Om_{ab}
\stackrel{\lea}{\d_c}i\hbar(-1)^{\ve_b+\ve_c}+{1\over
6}\eta^c\eta^b\eta^a[\Om_{abc},
\Om](-1)^{\ve_b+\ve_a\ve_c}+O(\eta^4).
\e{191}
To zeroth and first order in $\eta^a$ the master equation \r{18} is satisfied
identically. However, to second order in $\eta^a$ it yields exactly \r{13}.
 At  third order in $\eta^a$ it yields
\be
&&\d_a\Om_{bc}(-1)^{\ve_a\ve_c}+\half(i\hbar)^{-2}(\Om_a,
\Om_{bc})_{\Om}(-1)^{\ve_a\ve_c}+cycle(a,b,c)=\nn\\&&=-(i\hbar)^{-3}(\Om_a,
\Om_b,
\Om_c)_{\Om}(-1)^{\ve_a\ve_c}-{2\over3}(i\hbar)^{-1}[\Om'_{abc},\Om],\nn\\
&&
\Om'_{abc}\equiv\Om_{abc}-{1\over8}\left\{(i\hbar)^{-1}[\Om_{ab},
\Om_c](-1)^{\ve_a\ve_c}+cycle(a,b,c)\right\},
\e{20}
where
we have introduced a higher quantum antibracket defined by (this expression
may be
obtained from appendix B in \cite{Quanti})
\be
&&(A,
B,
C)_{Q}(-1)^{(\ve_A+1)(\ve_C+1)}\equiv\nn\\&&\equiv{1\over3}\left(  [(A, B)_{Q},
C](-1)^{\ve_C+(\ve_A+1)(\ve_C+1)}+cycle(A,B,C)
\right), \quad Q^2=0,
\e{21}
where the quantum antibracket on the right-hand side is given by \r{14}. It
satisfies Leibniz' rule if \r{14} does. The quantum antibracket \r{14} satisfies
the Jacobi identities only if
$(A, B,
C)_{Q}=0$ since we have
\be
&&(A, (B, C)_{Q})_{Q}
(-1)^{(\ve_A+1)(\ve_C+1)}+cycle(A,B,C)=\nn\\&&=-\half[(A,B,C)_{Q}(-1)^{(\ve_
A+1)(\ve_C+1)},
Q]
\e{211}
(cf. eq.(36) in \cite{Quanti}).

Comparing equation \r{20} and the integrability conditions of \r{13} we
find exact
agreement. We have also checked that the consistency condition
\r{19}  yields exactly
\r{9} to second order in $\eta^a$, which is consistent with \r{13} as it should.
Similarly we have checked that \r{19} to third
order in $\eta^a$  yields a condition which is consistent with \r{20},
exactly like
\r{9} is consistent with \r{13}.

 The
master equation \r{18} yields at higher orders in $\eta^a$ equations
 involving still higher quantum
$\Om$-antibrackets defined in terms of lower antibrackets like in \r{21}, and
operators
$\Om_{abc\ldots}$ with still more indices. We conjecture that these equations
agree exactly  with the integrability conditions of
\r{20}. For a rank-$N$ theory we expect that there exists a solution of the form
\r{16} to the master equation \r{18}, which terminates just at the maximal
order $\eta^{N}$ (for a particular example see next section).

\section{Example: Quasigroup first rank theories.}
As an illustration of our formulas we consider now constraint operators
$\theta_a$
forming a rank one theory in which case we have
\be
&&\Om=\cC^a \theta'_a+\half \cC^b\cC^a
U^c_{ab}\pet_c(-1)^{\ve_c+\ve_b},\quad\theta'_a\equiv\theta_a+\half i\hbar
U_{ab}^b(-1)^{\ve_b}.
\e{22}
The nilpotence of $\Om$ requires \r{6} and
\be
&&\left(i\hbar U^d_{ab}U^e_{dc}+[U^e_{ab},
\theta'_c](-1)^{\ve_c\ve_e}\right)(-1)^{\ve_a\ve_c}+{ cycle}(a,b,c)\equiv 0,
\e{23}
and in addition also restrictions on the commutators $[U^c_{ab}, U^f_{de}]$. The
latter condition is satisfied if
\be
&&[U^c_{ab}, U^f_{de}]=0, \quad [[\theta_d,
U^c_{ab}], U^g_{ef}]=0,
\e{231}
which corresponds to quasigroups \cite{Bat}.
 In this case $\Om_a$ may be chosen to be
\be
&&\Om_a(\phi)=\la^b_a(\phi)\pet_b,\quad \la^b_a(0)=\del^b_a,
\e{24}
where we assume that
\be
&&[\la^b_a, \la^d_c]=0 \quad\Ra\quad[\Om_a, \Om_b]=0.
\e{25}
The quantum antibracket \r{14} is then given by
\be
&&(\Om_a, \Om_b)_{\Om}=[\Om_a, [\Om, \Om_b]]=-(i\hbar)^2\la_a^f\la_b^e
U_{ef}^d\pet_d(-1)^{\ve_d+\ve_e+\ve_f+\ve_b\ve_f}+\nn\\&&+i\hbar
\left(\la^c_a[\theta'_c, \la^d_b ]-\la^c_b[\theta'_c,
\la^d_a](-1)^{\ve_a\ve_b}\right)\pet_d(-1)^{\ve_c}-\nn\\&&
-i\hbar\left(
\la^f_a\cC^e[U^c_{ef},
\la^d_b](-1)^{\ve_b(\ve_c+1)}-\la^f_b\cC^e[U^c_{ef},
\la^d_a](-1)^{\ve_a(\ve_c+1)+\ve_a\ve_b}\right)\pet_d\pet_c-\nn\\&&-
\cC^e[[\theta'_e, \la_b^c], \la^d_a
]\pet_d\pet_c(-1)^{(\ve_a+1)(\ve_b+\ve_c+1)}-
\nn\\&&-\half \cC^f\cC^e[[U^c_{ef},
\la^d_b], \la^g_a]\pet_g\pet_d\pet_c
(-1)^{\ve_c+\ve_f+(\ve_a+1)(\ve_b+\ve_c+\ve_d)+(\ve_b+1)(\ve_c+1)}.
\e{26}
If we also require
\be
&&(\Om_a, \Om_b, \Om_c)_{\Om}=0\quad \Lra \quad [(\Om_a, \Om_b)_{\Om}, \Om_c]=0,
\e{27}
then $(\Om_a, \Om_b)_{\Om}$ in \r{26} satisfies the Jacobi identities which
makes
\r{13} integrable if $\Om_{ab}=0$. This condition is satisfied if we impose
\be
&&[\la^b_a,
U^c_{de}]=0,
\quad [\la^b_a,[\la^d_c,
\theta_e]]=0.
\e{28}
Eq.\r{13} may now be written as
\be
&&\d_a\tilde{\la}_b^c-\d_b\tilde{\la}_a^c(-1)^{\ve_a\ve_b}=\tilde{\la}^e_a
\tilde{\la}^d_b
\tilde{U}^c_{de}(-1)^{\ve_b\ve_e+\ve_c+\ve_d+\ve_e}.
\e{29}
where $\tilde{\la}_a^b\equiv V\la_a^bV^{-1}$ and $\tilde{U}_{ab}^c \equiv
VU_{ab}^cV^{-1}$ where in turn the operator $V(\phi)$ is determined by the
equation
\be
&&i \hbar \d_a V = V \lambda^b_a \theta'_b (-1)^{\ve_b}.
\e{291}
 Eq.\r{27} and
$\Om_{ab}=0$ make all higher integrability conditions identically zero.
One may note that
\be
&&Y_a(\phi)=(i\hbar)^{-1}[\Om, \Om_a]=\la^b_a
\theta'_b(-1)^{\ve_a+\ve_b}+\nn\\&&+
\la^b_a\cC^dU^c_{db}\pet_c(-1)^{\ve_a+\ve_c}+(i\hbar)^{-1}\cC^b[\theta_b,
\la^c_a]\pet_c.
\e{30}

\section{Application: Generating operators for quantum antibrackets.}
In \cite{Quanti} it was shown that quantum antibrackets for commuting
operators or
operators satisfying a nonabelian Lie algebra may be derived from a generating
operator $Q(\phi)$. Here we generalize this construction to operators in
arbitrary
involutions. In distinction to the case in \cite{Quanti} we define here the
generalized generating operator
$Q(\phi)$ by equation \r{8}, \ie
\be
&&Q(\phi)\stackrel{\lea}{\nabla}_a\equiv
Q(\phi)\stackrel{\lea}{\d_a}-(i\hbar)^{-1}[Q(\phi), Y_a(\phi)]=0
\e{31}
with the boundary condition
\be
&&Q(0)=Q,\quad Q^2=0.
\e{32}
Since \r{31} implies
\be
&&[Q(\phi), Q(\phi)]\stackrel{\lea}{\d_a}=-(i\hbar)^{-1}[Y_a(\phi), [Q(\phi),
Q(\phi)]],
\e{33}
the boundary condition \r{32} implies $Q(\phi)^2=0$ which
shows that $Q\,\ra\, Q(\phi)$ is a unitary transformation.

Following ref.\cite{Quanti} we   define generalized quantum antibrackets in
terms
of
$Q(\phi)$ according to the formula
\be
&&(Y_{a_1},
Y_{a_2},\ldots,Y_{a_n})'_{Q(\phi)}\equiv-Q(\phi)
\stackrel{\lea}{\d}_{a_1}\stackrel{\lea}{\d}_{a_2}\cdots
\stackrel{\lea}{\d}_{a_n}(i\hbar)^{n}(-1)^{E_n},\nn\\&& E_n\equiv
\sum_{k=0}^{\left[{n-1\over 2}\right]}\ve_{a_{2k+1}}.
\e{34}
In particular we have then (At $\phi^a=0$ we have $Y_a(0)=\theta'_a+
\{\mbox{\small possible ghost
dependent terms}\}$.)
\be
&&(Y_a(\phi),
Y_b(\phi))'_{Q(\phi)}\equiv-Q(\phi)\stackrel{\lea}{\d_a}
\stackrel{\lea}{\d_b}(-1)^{\ve_a}(ih)^2=\nn\\
&&=\half\left([Y_a,
[Q(\phi), Y_b]]-[Y_b, [ Q(\phi),
Y_a]](-1)^{(\ve_a+1)(\ve_b+1)}\right)-\nn\\
&&-\half i\hbar[Q(\phi),
Y_a\stackrel{\lea}{\d_b}+Y_b\stackrel{\lea}{\d_a}(-1)^{\ve_a\ve_b}](-1)^{\ve_a},
\e{35}
where we have made use of \r{8}.
The third line, which makes \r{35}  differ from \r{14}, is the price of a
reparametrization independent extension of the definition \r{14} onto the
space of
parameters $\phi^{a}$. The reason for this is that \r{35} is a second
derivative of
a scalar, which is not a tensor. Thus only within a preferred coordinate
frame and
at least at a fixed value of the parameters $\phi^{a}$  can one expect the
formula
\r{35} to reproduce the original antibracket \r{14}. A similar situation
is naturally expected to occur for all higher
antibrackets as well. We also expect the canonical coordinates
which are well known in Lie group theory, to be  preferred ones
also in the general case, so that all antibrackets would be reproduced in these
coordinates at $\phi^{a} = 0$.

To illustrate the situation with the extension of the 3-antibracket, we give the
following explicit formula
\be
&&(Y_a(\phi), Y_b(\phi), Y_c(\phi))'_{Q(\phi)} (-1)^{(\ve_a + 1)(\ve_c +
1)} \equiv
 Q(\phi) \stackrel{\lea}{\d_a} \stackrel{\lea}{\d_b} \stackrel{\lea}{\d_c} (i
\hbar)^3 (-1)^{\ve_a
\ve_c} =\nn\\&& = {1\over 3} \left([(Y_a, Y_b)'_Q, Y_c] (-1)^{\ve_c + (\ve_a +
1)(\ve_c + 1)} + cycle (a, b, c)\right) +\nn\\&&+ {1\over 3} i \hbar \left([[Q,
Y_a], Y_b
\stackrel{\lea}{\d_c} + Y_c \stackrel{\lea}{\d_b} (-1)^ {\ve_b \ve_c}]
(-1)^{\ve_a
\ve_c} + cycle (a, b, c)\right) +\nn\\&&+ {1\over 6} (i
\hbar)^2 \left([Q, (Y_a \stackrel{\lea}{\d_b} + Y_b \stackrel{\lea}{\d_a}
(-1)^{\ve_a
\ve_b})
\stackrel{\lea}{\d_c}] (-1)^{\ve_a
\ve_c} + cycle (a, b, c)\right).
\e{36}
Obviously the $\hbar$-deviation from \r{21} vanishes together with the one
for the 2-antibracket in \r{35}. However, the vanishing of the
$\hbar^2$-deviation
imposes a new condition. In the corresponding $\phi$-extended $n$-antibrackets
obtained from \r{34} we expect to have deviations involving up to
$n-2$ cyclically symmetrized derivatives of the 2-antibracket deviation which in
terms of canonically coordinates should vanish at $\phi^a=0$. The
$\phi$-extended
$n$-antibrackets will then exactly  reproduce the
original $n$-antibrackets  at $\phi^a = 0$.

For first rank theories equation \r{31} reduces to
\be
&&Q(\phi)\stackrel{\lea}{\d_a}=(i\hbar)^{-1}[Q(\phi), \la^b_a(\phi)
\theta'_b(-1)^{\ve_a+\ve_b}]
\e{37}
if we assume that $Q(\phi)$ commutes with the last terms in \r{30}. Eq.\r{37}
coincides exactly with  equation (67)  in appendix B of \cite{Quanti} if
$U_{ab}^b(-1)^{\ve_b}=0$ or if $Q(\phi)$ commutes with $\la^b_a
U_{bc}^c(-1)^{\ve_c}$.
\\
\\

\noindent
{\bf Acknowledgments}

I.A.B. would like to thank Lars Brink for his very warm hospitality at the
Institute of Theoretical Physics, Chalmers and G\"oteborg University. The
work is partially supported by INTAS-RFBR grant 95-0829. The work of I.A.B.
is also
supported by INTAS grant 96-0308 and by RFBR grants 96-01-00482, 96-02-17314.
I.A.B. and R.M. are thankful to the Royal Swedish Academy of Sciences for
financial support.


\begin{thebibliography}{Simple}

\bibitem{Quanti}I. A. Batalin and
R. Marnelius,
 \ {\em Quantum antibrackets}, \ {\tt hep-th/9805084}\\(to be published in
Phys. Lett. B).

\bibitem{BFV}I. A. Batalin and G. A Vilkovisky, \ {\sl  Phys. Lett.}
\ {\bf B69}, 309 (1977),\\
 E. S. Fradkin and T. E. Fradkina, \ {\sl  Phys. Lett.}
\ {\bf B72}, 343 (1978),\\
I. A. Batalin and E. S. Fradkin, \ {\sl  Phys. Lett.}
\ {\bf B122}, 157 (1983).

\bibitem{BF}I. A. Batalin and E. S. Fradkin,\ {\sl Phys. Lett.}\ {\bf B128}, 303
(1983);\\ {\sl Riv. Nuovo Cim.}\ {\bf 9}, 1
(1986);\ {\sl Ann. Inst. Henri Poincar\'{e}}\
{\bf 49}, 145 (1988).

\bibitem{Bat}I. A. Batalin, \ {\sl  J. Math. Phys.}
\ {\bf 22}, 1837 (1981)

\end{thebibliography}
\end{document}